\documentstyle[12pt,epsf]{article}

\def \double  {\baselineskip=  2  \normalbaselineskip}
\newcommand{\beq}{\begin{equation}}
\newcommand{\eeq}{\end{equation}}
\newcommand{\beqa}{\begin{eqnarray}}
\newcommand{\eeqa}{\end{eqnarray}}
\newcommand{\beqan}{\begin{eqnarray*}}
\newcommand{\eeqan}{\end{eqnarray*}}
\newcommand{\no}{\nonumber}

\newcommand{\ol}{\overline}
\newcommand{\ra}{\rightarrow}

\newcommand{\ben}{\begin{enumerate}}
\newcommand{\een}{\end{enumerate}}
\newcommand{\bfl}{\begin{flushleft}}
\newcommand{\efl}{\end{flushleft}}
\newcommand{\ba}{\begin{array}}
\newcommand{\ea}{\end{array}}
\newcommand{\btab}{\begin{tabular}}
\newcommand{\etab}{\end{tabular}}
\newcommand{\bit}{\begin{itemize}}
\newcommand{\eit}{\end{itemize}}

\newcommand{\cO}{{\cal O}}

\newcommand{\vs}{\vspace}
\newcommand{\hs}{\hspace}

\def \5 {\gamma_{5}}

\def \G {\Gamma}
\newcommand{\prepr}[1] {\begin{flushright} {\bf #1} \end{flushright} \vskip
1.5cm}
\newcommand{\titul}[1] {\begin{center}{\Large {\bf #1 } } \end{center}\vskip
1.cm}

\newcommand{\autor}[1] {\begin{center} {\bf \lineskip .3cm #1  }
                        \end{center} }

\double
\newcounter{muni}

\begin{document}
\vspace{.1cm}
\hbadness=10000
\pagenumbering{arabic}
\begin{titlepage}
\prepr{Preprint hep-ph/mmmnnn\\PAR/LPTHE/95-50  \\06 November 1995 }
\titul{CP violation in $\eta  \ra \pi + \pi $ by Higgs
- $\eta$ mixing  \\ through two-loop quantum effects }
\autor{ X. Y. Pham \footnote{\rm Postal address: LPTHE, Tour 16, $1^{er}$
Etage, Universit\'e Pierre {\it \&} Marie Curie and Universit\'e Denis Diderot,
4 Place Jussieu, F-75252 Paris CEDEX 05, France.\\
Email address : pham@lpthe.jussieu.fr }}


\begin{center}
{\it Universit\'e  Pierre et Marie Curie, Paris VI \\
   Universit\'e Denis Diderot, Paris VII \\
 Physique Th\'eorique et Hautes \'Energies \\
Unit\'e associ\'ee au CNRS : D 0280
 }
\end{center}

\vs{-13cm}
\thispagestyle{empty}
\vs{120mm}
\noindent

\vs{10mm}
\begin{abstract}
Strictly within the standard electro-weak interaction , CP violation
in the flavour conserving process $\eta \ra \pi + \pi $ could originate
from the mixing of the $\eta$ meson with the virtual scalar Higgs $ H^{0}$
via $ W^{+}+W^{-}$ and $Z^{0}+Z^{0}$ exchange.

The parity-violation carried by these weak gauge bosons makes the mixing
possible at two-loop level. Nowhere the Kobayashi-Maskawa (KM) phase
mechanism is needed.

For the Higgs mass between 100-600 GeV ,the  $\eta \ra \pi +\pi $ branching
ratio is found to be $3\cdot 10^{-26} - 2\cdot 10^{-29}$ , hence unconventional
CP violation mechanisms are the only ones that could give rise to its
observation at the
existing or near future  $\eta$ factories, unless the Higgs mass is improbably
as
light as 10 MeV.

\end{abstract}

\vs{5mm}
\end{titlepage}

\newpage

To understand the origin and the nature of CP violation, in addition to
the studies of flavour changing K and B mesons processes, investigations are
also needed in flavour conserving ones \cite{R1} for which $\eta \ra \pi +  \pi
$
decay and electric dipole moment of baryons are some typical examples. Like
the $K_{L}^{0}$ , an eventual coexistence of both three and two-pion decay
modes
of the $\eta$ would imply that CP is violated in the flavour conserving sector.
Therefore experimental searches for the two-pion decay mode of the $\eta$ is of
great interest\cite{R1}, and the purpose of this letter is to give a reliable
estimate of
its branching ratio, strictly within the standard electro-weak Higgs framework.

We do find indeed that CP violation in  $\eta \ra \pi +  \pi $ simply can be
generated by the neutral Higgs boson, {\it independently of all other
mechanisms  }.

In principle, an eventual role played by the standard neutral Higgs boson
$ H^{0}$ in CP odd interaction should be envisaged at first, since it lies
in the heart of the standard model (SM).We even do not have to evoke the
KM phase. The latter would be, in some sense, the next step to be considered
in the studies of $\eta \ra \pi +\pi$ decay, while the third one could be non
conventional mechanisms of CP violation, for example the $\Theta$
vacuum\cite{R2}
in QCD or spontanous breaking triggered by charged scalar fields\cite{R3}.

With a touch of irony,the third stage - CP odd $\Theta$  term - as source
of $\eta \ra \pi + \pi $ decay was worked out long time ago\cite{R4} before the
second one for which the rate is recently computed by Jarlskog and Shabalin
\cite{R5}
( JS ) within the KM framework of the penguin effective lagrangian \cite{R6}.
As far as we know, the most straightforward implication of the standard Higgs
mechanism on $\eta \ra \pi + \pi $ reaction has never been noticed, although in
principle
it should be naturally considered as the first step to cross over.
The purpose of this work is to fulfil the gap: we indeed show that this mode
actually occurs because of the pseudoscalar-scalar (PS) mixing between
the  $\eta$  and the virtual neutral Higgs boson which subsequently
decays into two pions ; this PS mixing is due to quantum effects at two-loop
level as shown in Figure 1 .

Physically, it means that the parity-violation  VA  property of the weak bosons
shifts the intrinsic  CP = - 1 of the  $\eta$  into the  CP  = + 1  of the
$H^{0}$  : {\it parity-violation turns out to be the source of  CP non-
conservation,
due to the Higgs and gauge bosons interplay}. This observation is illustrated
by explicit computation of the diagram, the relevant quantity to be considered
is :
\beq
I(k^2) \equiv (-1) \int {d^4 q \over (2\pi)^4} {d^4 p \over (2\pi)^4}
{Trace \{ \gamma_{\mu} (a - b \5 )(\not{p} + m) \5 (\not{p} + \not{k} + m)
\gamma^{\mu} (a - b \5 ) (-\not{q} + m^{'}) \} \over
(q^2 - {m^{'}}^2)(p^2 - m^2)((p+k)^2 - m^2)((p+q)^2 - M^2)((p+q+k)^2 - M^2)}
\label{eq:1}
\eeq
with $a = b = V_{cs}$, $ m = m_{s}, m^{'} = m_{c}, M = M_{W} $ for  $ W$
exchange.\\
$a = {(-1 + 4 \hs{2mm} sin^{2}\theta_{W}/3 ) \over \sqrt{2} \hs{2mm} cos
\theta_{W} } $ ,
$b = { -1 \over \sqrt{2} \hs{2mm} cos \theta_{W} } $ , $ m = m^{'} = m_{s} , M
= M_{Z} $
for $ Z $ exchange .

In Eq.(1) we have taken, as an illustrative example,the contribution of the
$s\ol{s}$ component of the  $\eta$  to the loops.
We first integrate over $ d^4 p $  using the  $ x , y , z $   Feynman
parametrization, and
then over  $ d^4 q $  , the result is :
\beq
I(k^2) =  C \hs{2mm} {m \hs{1mm} a \hs{1mm} b\hs{1mm} \over 32 \hs{1mm} \pi^{4}
}
\hs{2mm} {k^{2} \over M^{2}} [ 1 + \cO \{ {k^{2} \over M^{2}}, {m^{2} \over
M^{2}}, {{m^{'}}^{2} \over M^{2} } \} ]
\label{eq:2}
\eeq
where :
\beq
C   \equiv \int ^1_0 dx \int ^x_0 dy \int ^y_0 dz  {(z(1-y) - y(1-x)) \over
(x-z)^{3} (1-x+z)^{2}}  =  1/4
\label{eq:3}
\eeq
The expression (2) we obtain for the two-loop integration is impressively
simple, because higher orders in ${k^{2} \over M^{2}}$  , ${m^{2}\over M^{2}}$
,
${{m^{'}}^{2}\over M^{2}}$ ( beyound the linear term ${k^{2} \over M^{2}}$ )
are
neglected in the course of our  $d^{4}q $  integration. In the  $d^{4}p$  one ,
everything is kept however. Without this legitimate approximation,we would
obtain
an avalanche of - unnecessary and numerically negligible- complicated
expressions
involving,among others,dilogarithmic (or Spence) function frequently met in
such circumstance.

It remains two questions to be settled : the first one concerns the effective
point-like coupling constant $ g_{\eta Q \ol{Q}} $ of the $\eta$ meson with
quarks Q assumed in Fig.1.  Is it justified ? If yes what is its numerical
values ?
The second point deals with the off-shell (virtually light mass $k^{2} =
m_{\eta}^{2}$ )
Higgs decay amplitude into two pions.

1- The justification for the $\eta$-quarks coupling can be traced back to the
famous
antecedent Adler, Bell, Jackiw ( ABJ ) chiral anomaly\cite{R7} inherent to
$\eta \ra \gamma +\gamma$ ( or $\pi^{0} \ra \gamma +\gamma $ ) decay for which
the same triangle one-loop is involved where external real photons
replace internal virtual gauge bosons of our two-loop diagram in Fig.1.
Such $\eta$ (or $\pi$ ) coupling to light quarks seems to stand on a firm
ground and
is intimately connected to its Goldstone nature,to the partially conserved
axial current ( PCAC ) and its consequence : the Goldberger-Treiman ( GT )
relation.
These properties hold\cite{R8,R9} also in the Nambu,Jona-Lasinio
model\cite{R9},
a prototype of low-energy effective lagrangian suitable for studying Goldstone
particles.

However there is an important difference between the external on-shell photons
in $\pi^{0} \ra \gamma +\gamma $  decay\footnote{For simplicity we take $\pi$
as an example
,the $\eta$ case is similar although the situation is getting complicated by
the SU(3)
flavour singlet-octet $\eta - \eta^{'}$ mixing laterly included} and the
internal off-shell weak gauge bosons considered in Fig.1 . For $\pi^{0} \ra
\gamma+\gamma$, in order to get the right answer in agreement with data and
with the ABJ anomaly i.e. the amplitude is proportional to  $ {g_{\pi Q
\ol{Q}}\over m_{Q}} =  {1 \over f_{\pi}} $ , two
conditions have to be satisfied :

(i) $k^{2} \ll 4 m_{Q}^{2} $,such that the triangle loop integration \cite{R11}
yields term proportional to $ {1\over m_{Q}}$  ($ m_{Q}$ in the { \it
denominator} ).

(ii) the validity of the GT relation at the quark level,i.e.
${ g_{\pi Q \ol{Q}} \over m_{Q}}$
 = ${1 \over f_{\pi}}$.

These two conditions are naturally fulfilled with the Goldstone nature of the
pion. The
case $k^{2} > 4m_{Q}^{2}$ ( not choosen by Nature ) would lead to catastrophic
disagreement with data and with the ABJ anomaly,since instead of $ {1 \over
m_{Q}}$
one would get\cite{R11}   -
$ {m_{Q} \over k^{2} }
[ \ ln {(\sqrt{k^{2}} + \sqrt{k^{2} - 4 m_{Q}^{2}}))
\over (\sqrt{k^{2}} - \sqrt{k^{2} - 4 m_{Q}^{2}})} - i \pi ]^{2}$ .

The independence of the ABJ anomaly on the  internal quark mass, a well known
fact,
has its origin in the  $k^{2} \ll 4 m_{Q}^{2}$
constraint\footnote{Incidently,the same condition is satisfied by the
Steinberger old calculation of $ \pi^{0} \ra \gamma+\gamma $ (Phys.Rev 76,
1180, (1949)) in which quark was proton at that time ; his work fits
naturally with the constituent u, d mass.}: the $\pi^{0} \ra \gamma +\gamma$
amplitude depends neither separately on $ g_{\pi Q \ol{Q}} $ nor on $m_{Q}$
but only on their ratio  $ {g_{\pi Q \ol{Q}} \over m_{Q}} $  which is fixed by
 $ {1 \over f_{\pi}}$.  As a consequence, the $\pi^{0} \ra \gamma +\gamma $
rate is
helpless for indicating which values of  $g_{\pi Q \ol{Q}}$  or   $ m_{Q}$  to
be employed : current or constituent mass of the light quarks ?

In our case with both internal off-shell gauge bosons,the loop integration no
longer
produces ${1 \over m_{Q}}$ but instead  $m_{Q}$ in the {\it numerator}, as
explicitly
shown in Eq.(2).
Our  PS  mixing is then proportional to
$ m_{Q} g_{\eta Q \ol{Q}} $  = $ {m_{Q}^{2} \over f_{\eta}} $
such that the choice of  $m_{Q}$ is unavoidable.
We have seen that the real photons rate cannot help, nevertheless the
$ k^{2} \ll 4 m_{Q}^{2}$ lesson must not be forgotten. This condition
favourably hints to the choice of constituent mass, as also found by other
authors \cite{R9,R12} in a different context.

   2- The virtually light ($ k^{2} = m_{\eta}^{2}$ ) Higgs boson coupling to
two
pions can be reliably estimated from the so-called conformal anomaly\cite{R13}
i.e. the trace of the energy- momentum tensor in QCD\cite{R13}
$\Theta^{\mu}_{\mu} = - \beta_{0} { \alpha_s \over 8 \pi } G_{\mu\nu}
G^{\mu\nu} $
$(\beta_{0} = 9 $ is the first coefficient of the QCD $\beta$ function ).
The crucial point-as explained in \cite{R14} - is that the matrix element of
the operator
$\alpha_s G_{\mu\nu} G^{\mu\nu} $
between the two-pion state and vacuum is nonvanishing in the chiral limit, it
even does
not depend on $\alpha_{s}$ ; the (virtually light $k^{2}$) Higgs decay
amplitude into two pions is found to be\cite{R14} :
\beq
f_{H\pi\pi}(k^2) = - {g \over \beta_{0}} \hs{2mm}
{k^2 + 5.5 \hs{2mm} m_{\pi}^2 \over M_{W} } \label{eq:4}
\eeq
where $ g = e /sin\theta_{W}  $ is the standard SU(2) gauge coupling
which enters also
in the three other vertices of Fig.1.
Putting altogether the ingredients, we obtain for the $\eta \ra \pi +\pi$
decay amplitude the following result :
\beq
A_{\eta\pi^{+}\pi^{-}} = A_{\eta\pi^0\pi^0} =
{1 \over 6 \sqrt{3}} \hs{2mm}
\left( {G_F M_W^2 \over 4 \pi^2} \right)^2 \hs{2mm}
{m_{\eta}^2 \over M_W^2} \hs{2mm}
{ m_{\eta}^2 + 5.5 m_{\pi}^2 \over m_{H}^2} \hs{2mm}
{1 \over f_{\eta}} \hs{2mm}
\left( X_W + { X_Z \over 4 cos^2\theta_{W}} \right) \label{eq:5}
\eeq
with
\beqa
X_W &=& m_s^2 (\sqrt{2} cos\theta_P + sin\theta_P) - (m_u^2 + m_d^2)
\left( { cos\theta_P \over \sqrt{2}}  - sin\theta_P \right) \label{eq:6}
\cr
X_Z &=& \left( 1 - {4 \over 3} sin^2\theta_W \right) \hs{2mm}
\left[ m_s^2 (\sqrt{2} cos\theta_P + sin\theta_P) -  m_d^2
\left( { cos\theta_P \over \sqrt{2}}  - sin\theta_P \right) \right] \no \\
\cr
& & \hs{20mm} -
\left( 1 - {8 \over 3} sin^2\theta_W \right)
m_u^2
\left( { cos\theta_P \over \sqrt{2}}  - sin\theta_P \right)  \label{eq:7}
\eeqa
from which :
\beq
\G(\eta \ra \pi^{+}\pi^{-}) = 2 \hs{2mm} \G(\eta \ra \pi^0 \pi^0) =
{|A_{\eta\pi\pi}|^2 \over 16 \pi m_{\eta}} \hs{2mm}
\sqrt{1 -{4 m_{\pi}^2 \over m_{\eta}^2} } \label{eq:8}
\eeq

In Eqs. (5) -(6),the GT like relation $g_{\eta Q \ol{Q}}  =  {m_{Q} \over
f_{\eta}}$
is used, quark color indices are summed up, and  $\theta_{P} \simeq -19^{o}$
is
the flavour SU(3) $ \eta -\eta^{'} $ mixing angle determined from their two
photon
rates. We take  $f_{\eta} = f_{\pi} \simeq 93$ $ MeV$.
Surprisingly enough ,it turns out that the numerical values of the quantity
$ Y \equiv  {(X_{W} + (X_{Z} /4\cos^{2}\theta_{W})) \over f_{\eta}} $ entering
in Eq. (5) is relatively insensitive to the choices of quark
masses : for the constituent ones $ m_{s} = 500$ $ MeV$, $m_{u} = m_{d} = 300$
$ MeV$,we have
$Y = 1.02$ $ GeV $ ; for the current ones
$m_{s} = 200$ $ MeV$, $m_{u} = m_{d} = 8$ $ MeV$,we get $Y = 0.52 $ $GeV$.
With the constituent mass choice ,we obtain : $Br(\eta \ra \pi + \pi ) =
3.10^{-26} \hs{2mm} ( {100 GeV \over M_H })^{4} $
such that for the Higgs mass
between 100 $GeV$ and 600 $GeV$,
the branching ratio into both charged and neutral pions
of the $\eta$ meson  varies in the range $ 3 \cdot 10^{-26}  -  2 \cdot
10^{-29} $
,which is similar although (for the Higgs mass $\leq 250 GeV$ ) somewhat larger
than the JS result.

Therefore the standard model predicts that existing as well as future $\eta$
factories
(Saturne, Celsius, Daphne ) could not detect the $ \eta \ra \pi +\pi $ mode
(unless
the Higgs mass is improbably as light as 10 $MeV$), implying that {\it
unconventional}
CP violation mechanisms are the only ones that could give rise to its eventual
observation.
This conclusion is not as negative as it seems, since as noted by JS, New CP
violation
mechanisms, what ever they may be, will have a golden opportunity to show up in
the $ \eta \ra \pi +\pi$ decay.

\vs{5mm}
\hspace{1cm} \large{} {\bf Acknowledgements}    \vspace{0.4cm}
\normalsize

It is a pleasure to thank M. Gourdin, Q.Ho-Kim and Y. Y. Keum for helpful
discussions.


\vs{55mm}
\hs{10mm}
\large
Figure Caption  :  \normalsize
\vs{5mm}

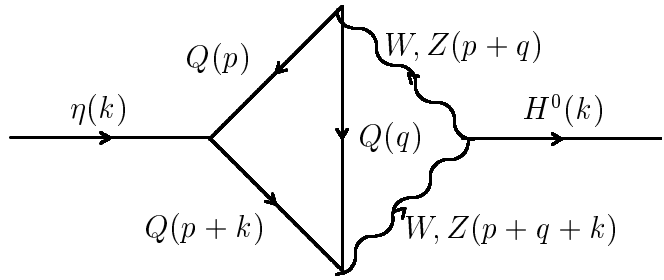
Figure 1 : $\eta $ - Higgs  mixing by two-loop quarks-gauge bosons exchange


\begin{thebibliography}{99}
\bibitem{R1} I.Yu Kobzarev and L.B. Okun , JETP{\bf 19} ,958 (1964).
%
\bibitem{R2} G. {'t} Hooft, Phys.Rev. Lett. {\bf 37}, 8 (1976)\\
A review is given by R.D. Peccei : Strong CP problem,
edited by C. Jarlskog in CP violation (Advanced series on directions in high
energy physics, vol 3 ,World Scientific ( 1989 ).
%
\bibitem{R3} T.D. Lee, Phys.Rev. {\bf D8}, 1226 (1973) \\
S.Weinberg, Phys. Rev. Lett. {\bf 31},657 (1976) \\
G.C.Branco,Phys.Rev. {\bf D22},2901 (1980) \\
A review is given by I.I.Bigi, A.I. Sanda and N.G. Uraltsev :Addressing  the
Mysterious with the Obscure,edited by G. Kane in Perspectives on Higgs Physics
(Advanced series on directions in high energy physics, vol 13 ,World Scientific
( 1993).
%
\bibitem{R4}
M.A.Shifman,  A.I.Vainstein and V.I.Zacharov, Nucl.Phys.{\bf B166}, 494 (1980)
\\
R.J.Crewther, P.Di Vecchia, G.Veneziano and E.Witten , Phys.Lett.{\bf B88},123
(1979),
{\bf B91}, 487 (E) (1980)
%
\bibitem{R5} C.Jarlskog and E.Shabalin ,  Phys. Rev.{\bf D52}, 248 (1995).
%
\bibitem{R6} A.I. Vainstein,V.I.Zacharov and M.A.Shifman, JETP{\bf 451},670
(1977).
%
\bibitem{R7} S.L.Adler, Phys.Rev.{\bf 177},2426 (1969) \\
J.S.Bell and R.Jackiw,Nuovo Cimento {\bf 60A},47 (1969).
%
\bibitem{R8} S.P.Klevansky , Rev.Mod.Phys.{\bf 64},649 (1992).
%
\bibitem{R9} M.Takizawa and M.Oka, Phys.Lett.{\bf B359},210 (1995).
%
\bibitem{R10} Y.Nambu and G.Jona-Lasinio, Phys.Rev.{\bf 122},345 (1961).
%
\bibitem{R11} See some textbooks , for example B.Dewitt and J.Smith:Field
theory in Particle physics p.284,North.Holland Publ. 1986.
%
\bibitem{R12} Y.Y.Keum and X. Y.Pham , Mod.Phys.Lett.{\bf A9}, 1545 (1994).
%
\bibitem{R13}R.J.Crewther, Phys.Rev.Lett.{\bf 28},1421 (1972) \\
M.Chanowitz and J.Ellis, Phys.Lett{\bf B40},397 (1972) \\
J.Collins,L.Duncan and S.Joglekar, Phys.Rev{\bf D16},438 (1977).
%
\bibitem{R14} M.B.Voloshin, Sov.Journ.Nucl.Phys.{\bf 44},478 (1987) \\
M.B.Voloshin and V.Zacharov, Phys.Rev.Lett.{\bf 45},688 (1980).


\end{thebibliography}
\end{document}